\documentclass[%
reprint,
superscriptaddress,
 amsmath,amssymb,
 aps, physrev,
]{revtex4-2}

\usepackage{xcolor}
\usepackage[section]{placeins}
\usepackage{graphicx}
\usepackage{dcolumn}
\usepackage{bm}
\usepackage[mathlines]{lineno}
\usepackage{comment}

\usepackage[colorlinks]{hyperref}
\hypersetup{%
    colorlinks=true
    ,urlcolor=blue
    ,citecolor=blue
    ,linkcolor=blue
}
\usepackage{booktabs}
\usepackage{makecell}

\begin{document}

\title{\textbf{Optimal-Transport Contrastive Learning for Centrality Determination in Heavy-Ion Collisions}}

\author{Shu-Yun Yang}
\email{yang@glmc.edu.cn}
\affiliation{School of Artificial Intelligence in Medicine, Guilin Medical University, Guilin, 541199, China}
\author{Yu Zhang}
\email{zhangyu@gxnu.edu.cn}
\affiliation{Guangxi Key Laboratory of Nuclear Physics and Technology, Guangxi Normal University, Guilin 541004, China}

\date{\today}

\begin{abstract}
	Determining centrality of heavy-ion collisions with high resolution is essential for event-by-event analyses, in particular for net-proton fluctuation measurements aimed at locating the QCD critical point. Supervised machine-learning approaches rely on labels generated by model and therefore inherit the model dependence of the training target. In this work, we propose an unsupervised method to determine collision centrality based on optimal transport. Each collision is represented by its 1-Wasserstein distances to a fixed set of landmark events chosen to cover the measured multiplicity range. A point-cloud network trained by contrastive learning then compresses these distances into a compact event embedding, which is partitioned into equal-size centrality classes. We find that this approach groups events with similar $N_{\rm part}$ and yields a better centrality resolution, quantified by the normalized width of the $N_{\rm part}$ distribution, than the conventional method of cutting on particle multiplicity. The method, however, encounters a bottleneck once realistic kinematic cuts are applied to the particles. We attribute this degraded performance to the loss of distinct features between central and peripheral collisions because of applying kinematic cuts. The application of optimal transport to centrality determination in experimental data, remains an open problem that needs further study.
\end{abstract}

\maketitle

\section{\label{sec:intro}Introduction}

One of the main goals of heavy-ion collisions is to explore the phase structure of strongly interacting matter. At extremely high temperature and baryon density a deconfined quark-gluon plasma (QGP) is expected to be created. Lattice QCD calculations show that at vanishing baryon chemical potential ($\mu_B$) the transition between the QGP and the hadronic phase is a smooth crossover~\cite{Aoki:2006we}, while at large $\mu_B$ the fermion sign problem prevents direct lattice calculations and effective models indicate that the transition becomes of first order. The end point of the first-order boundary is the QCD critical end point (CEP). Considerable theoretical and experimental effort has been devoted to locating it; recent theoretical work converges on the high-baryon-density region around $\mu_B\sim 600$ MeV~\cite{Fu:2023lcm,Lu:2025cls,Cai:2022omk,Hippert:2023bel,Basar:2023nkp,Shah:2024img}, which is accessible to the RHIC Beam Energy Scan program~\cite{STAR:2017sal,Chen:2024aom} and the STAR fixed-target experiment~\cite{Meehan:2016iyt}.

Fluctuations of conserved quantities such as net baryon number, net electric charge, and net strangeness are proposed as sensitive observables for the QCD critical point~\cite{Stephanov:2008qz,Luo:2017faz,Pandav:2022xxx}. The analysis, however, is complicated by initial-volume fluctuations~\cite{Skokov:2012ds}: events with the same impact parameter or the same number of participants can still have different volumes, and this irreducible fluctuation limits the precision of event-by-event measurements.

Centrality, the degree of overlap of the two colliding nuclei, is the key input of all such analyses. It can be quantified by the impact parameter $b$, by the number of participating nucleons $N_{\rm part}$ (those that experienced at least one inelastic collision), or by the number of binary collisions $N_{\rm coll}$. In experiment, centrality is determined from a final-state reference quantity, usually the charged-particle multiplicity $N_{\rm ch}$: the measured $N_{\rm ch}$ distribution is compared with that of a Monte Carlo model~\cite{Miller:2007ri} and divided into percentiles, yielding classes such as 0--5\%. The reference quantity is not in one-to-one correspondence with the collision geometry, so a fixed $N_{\rm ch}$ interval corresponds to a distribution of $b$ and $N_{\rm part}$; the width of this distribution defines the centrality resolution. Even with an ideal reference an irreducible spread remains because of initial-volume fluctuations and of the finite number of particles. On the experimental side, specifically for event-by-event net-proton fluctuation measurements, several methods~\cite{Skokov:2012ds, Luo:2013bmi, Braun-Munzinger:2016yjz, Holzmann:2024wyd, Wang:2025fve} have been proposed and applied in data analysis to improve centrality resolution and reduce the volume-fluctuation effects associated with the use of a centrality reference.

Machine learning has been used for centrality determination for three decades. In 1994 a feedforward neural network was first applied to heavy-ion collisions, predicting the impact parameter as a supervised regression task~\cite{Bass1994}. In 1995 a combination of three observables was shown to determine the impact parameter about four times more accurately than any single one of them, and all single-observable methods were found to fail for central collisions~\cite{David1995}. The 1996 paper established the paradigm by discarding pre-processed observables and feeding the raw $(p_{\rm T}, p_{\rm z})$ momentum distribution of all outgoing particles directly into a two-layer perceptron~\cite{Bass1996}. The network performance, the average difference between reconstructed and true impact parameter, are claimed improved by a factor of 2 than classical techniques based on particle multiplicity, $p_{\rm T}$, etc. The authors also warned that training on a transport model introduces a model bias, proposing a cross-model test to estimate its magnitude.  In 2016, authors in Ref.~\cite{Altsybeev2017} combined information from several detector subsystems simultaneously, and performed both regression and classification tasks in centrality determination. Both centrality resolution and purity of centrality classes are improved by machine learning based techniques.
Machine learning development in centrality determination then moved to raw
detector-level input --- a point-cloud network applied directly to CBM hits and reconstructed tracks from full detector simulation~\cite{OmanaKuttan2020} and to higher accuracy on simulated data, where a CNN combined with LightGBM~\cite{Li2020}.
Similar works using CNN are also seen in Ref.~\cite{Murali2024, Wang2024}. There are also attempt using transfer learning technique, by fine-turning the pre-trained model to predict $N_{\rm part}$~\cite{Basak2025}.

Bayesian methods have also been used to reconstruct the impact-parameter distribution without modeling the collision dynamics, reaching a precision of about 5\% in Ref.~\cite{Das2018} and about 10\% in Ref.~\cite{Rogly2018}. The model dependence of the labeled training target itself has been investigated in Refs.~\cite{OmanaKuttan2023,Yue2026}. Unsupervised machine-learning methods, which do not depend on model-generated labels, are increasingly used in heavy-ion physics, but their application to centrality determination remains limited. Unsupervised machine learning methods are also used in heavy ion physics~\cite{Metodiev:2017vrx, Mikuni:2020qds,Thaprasop:2020mzp}, while few are for centrality determination.

In this work we use an unsupervised method to order events by centrality. Instead of a single reference quantity, we compare the full final-state momentum distributions of events by means of the optimal-transport (Wasserstein) distance. Optimal transport (OT) has recently been introduced in collider physics as a metric on jets and on events~\cite{Komiske2019,Komiske2020}, but its application to centrality determination has so far been limited. Because no model-generated $N_{\rm part}$ or $b$ label enters the training, the model dependence of supervised approaches is avoided at the training stage, and the embedding can in principle be trained on experimental data themselves. We tested the method using UrQMD~\cite{Bass1998,Bleicher1999} Au+Au events at $\sqrt{s_{\rm NN}}$ = 3.0, 5.2, and 19.6 GeV, which correspond to the STAR fixed-target and collider configurations, and compare the results with the conventional $N_{\rm ch}$ slicing.

This paper is organized as follows. Section~\ref{sec:method} discuss the methodology. Section~\ref{sec:result} shows results. Section~\ref{sec:summary} summarizes the paper.

\section{\label{sec:method}Method}

\begin{figure}[!htpb]
	\centering
	\includegraphics[width=1.0\linewidth]{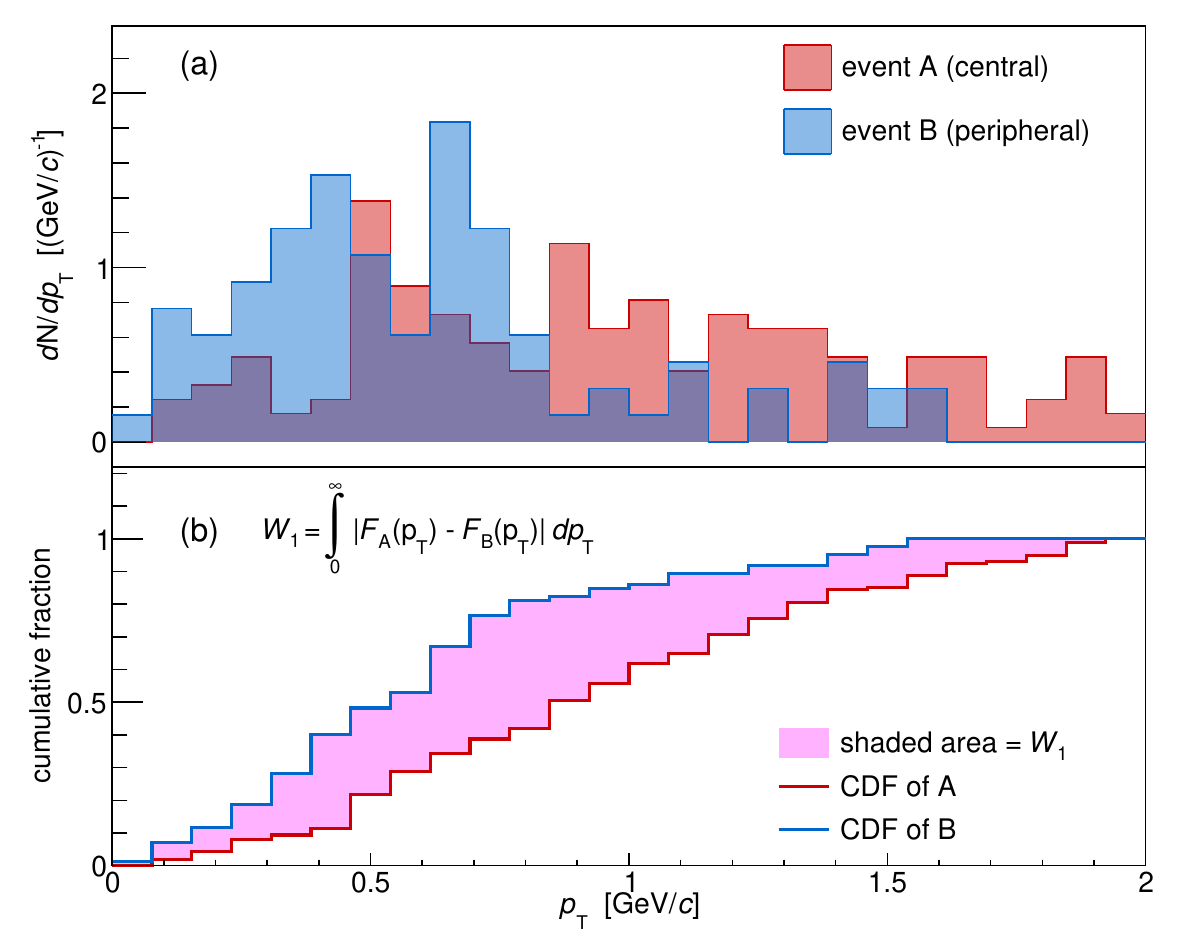}
	\caption{Panel (a): The $p_{\rm T}$ distributions of events with different centrality classes. The distributions are generated randomly for illustration. Panel (b): Cumulative distributions of event A and event B. The one dimensional Wasserstein distance, the least cost of transforming one distribution into another, is calculated by the shaded area enclosed by the two cumulative distributions.}
	\label{fig1}
\end{figure}

\subsection{\label{sec:w1}The Wasserstein distance}

An event is characterized by the three-momenta $\mathbf{p}_i=(p_{{\rm x},i},p_{{\rm y},i},p_{{\rm z},i})$ of its outgoing particles, i.e., by a cloud of $N_{\rm A}$ points in three-momentum space. To compare two events we use the optimal-transport (Wasserstein) distance, defined as the minimum average distance by which the particles of one event are moved to match those of the other; when the two events have different multiplicities, only $\min(N_{\rm A},N_{\rm B})$ pairs are matched and the surplus particles are left in place. For a particle $i$ of the first event and a particle $j$ of the second event this transport costs the Euclidean distance $d_{ij}=\lVert\mathbf{p}_i-\mathbf{q}_j\rVert$, and the distance between the two events is the average cost of the optimal one-to-one assignment,
\begin{equation}
	W_{1}=\frac{1}{\min(N_{\rm A},N_{\rm B})}\;\min_{\mathcal{M}}\sum_{(i,j)\in\mathcal{M}} d_{ij},
	\label{eq:w1}
\end{equation}
where the minimum runs over all one-to-one assignments $\mathcal{M}$ of $\min(N_{\rm A},N_{\rm B})$ particle pairs. When the two events have equal multiplicity, Eq.~(\ref{eq:w1}) is the 1-Wasserstein distance between their momentum distributions; in one dimension it reduces to the area enclosed by the two cumulative distributions, as illustrated in Fig.~\ref{fig1}. There is no closed-form expression for multi-dimensional momentum distributions, so Eq.~(\ref{eq:w1}) is evaluated numerically. We solve the assignment exactly with the Hungarian algorithm, implemented as the \texttt{linear\_sum\_assignment} routine of SciPy~\cite{Virtanen2020}; for equal multiplicities our estimate agrees with the exact optimal-transport solver of the POT library~\cite{Flamary2021}. When $N_{\rm A}\neq N_{\rm B}$, the surplus particles of the larger event are left unmatched at zero cost --- a partial-transport convention --- and the average in Eq.~(\ref{eq:w1}) runs only over the matched pairs. This choice deliberately removes the multiplicity difference from the distance, so that the optimal-transport features characterize the shape of the momentum cloud rather than the number of particles.
\begin{table}[!htpb]
	\centering
	\begin{tabular}{@{}c|c@{}}
		\toprule
		notation                     & \makecell[c]{Method to determine  centrality classes}                           \\ \midrule
		OT, all                      & \makecell[c]{OT-contrastive learning \\using all particles}                     \\ \midrule
		OT, charged                  & \makecell[c]{OT-contrastive learning \\using charged particles}                 \\ \midrule
		OT, charged, $\eta$          & \makecell[c]{OT-contrastive learning \\using charged particles with $\eta$ cut} \\ \midrule
		$N_{\rm ch}$ slicing         & Cutting $N_{\rm ch}$ distribution                                               \\ \midrule
		$N_{\rm ch}$ slicing, $\eta$ & \makecell[c]{Cutting $N_{\rm ch}$ distribution with $\eta$ cut}                 \\ \bottomrule
	\end{tabular}
	\caption{The notations of results shown in Sec.~\ref{sec:result}. Specifically, the $\eta$ cut is applied to charged particles with $-2.4 < \eta_{\rm lab} < 0$ in the laboratory frame for $\sqrt{s_{\rm NN}}$ = 3.0 and 5.2 GeV, and $|\eta| < 1$ in the center-of-mass frame for $\sqrt{s_{\rm NN}}$ = 19.6 GeV, which are the kinematic cuts used in STAR fixed-target and collider data analysis, respectively.}
	\label{tab:notation}
\end{table}

The ground metric is the Euclidean distance between three-momenta in a fixed frame. Shown in top three rows of Tab.~\ref{tab:notation}, we calculated $W_{1}$ for three scenarios in this work using: all outgoing particles, only charged particles, and charged particles with a $\eta$ cut corresponding to STAR fixed-target and collider data acceptance. For the scenario that use all particles, the momenta are taken in the center-of-mass frame; for the fixed-target energies the laboratory frame is obtained by a longitudinal boost of the stored center-of-mass momenta. For the scenarios with a $\eta$ cut, only charged particles inside the acceptance window are kept and their momenta are expressed in the same frame.

Computing $W_{1}$ between all pairs of events is not realistic: for $10^{6}$ events one needs $10^{12}$ transport problems, and modern experiments record tens to hundreds of millions of events. We therefore keep a fixed set of $N_{L}=100$ landmark events and represent each event by its distances to the landmarks, $F_{j}=W_{1}(\text{event},\text{landmark}_{j})$. The landmarks are chosen from the training sample so that their $N_{\rm ch}$ values are uniformly sampled across the event sample, which ensures that the feature basis covers the whole measured multiplicity range; replacing $N_{\rm ch}$ by $N_{\rm part}$ was checked to have no significant effect. Each of the 100 features is standardized over the training sample. This reduces the number of transport problems from $O(N^{2})$ to $O(NN_{L})$ and, more importantly, turns each event into a fixed-length vector that can be processed by a neural network.

Even $NN_{L}$ transport problems are too many for the production samples: computing the 100 distances for 7 million events would require $7\times10^{8}$ optimal-transport solutions. We therefore evaluate the feature vector $F$ only for 2000 training events and learn a map from these events that can be applied to the full event sample.

\begin{figure}[!htpb]
	\centering
	\includegraphics[width=1.0\linewidth]{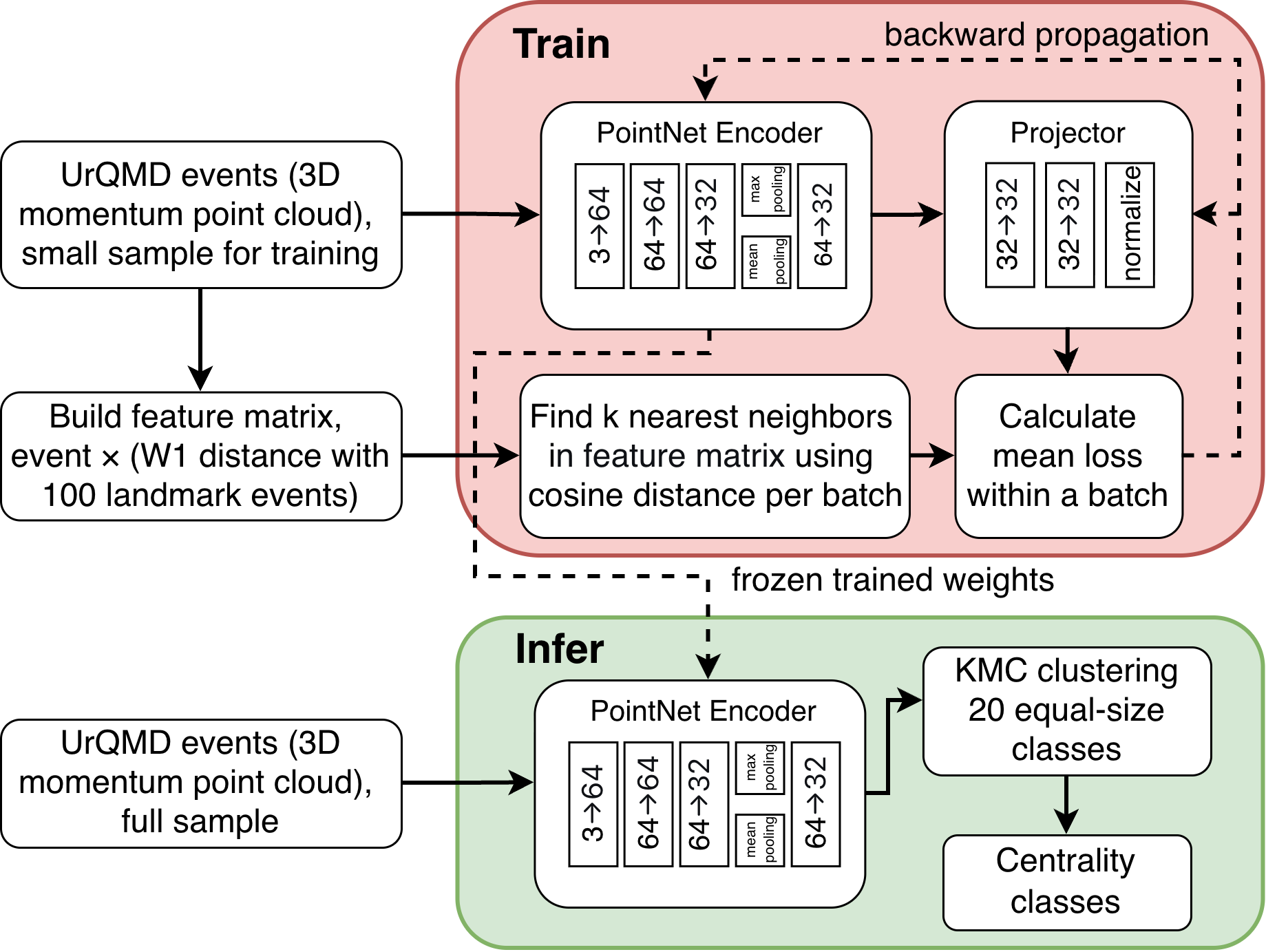}
	\caption{The workflow of the proposed method using optimal transport and contrastive learning. A small amount of UrQMD events are used to calculate Wasserstein distance and to train a PointNet network with contrastive learning method. The remaining events are used to infer their centrality classes with trained network.}
	\label{fig2}
\end{figure}
\subsection{\label{sec:distill}OT-Contrastive learning}
We introduce the proposed method named OT-contrastive learning in this section. The workflow is summarized in Fig.~\ref{fig2}.
The map from the raw particle cloud to a compact embedding, event-level feature vector, is a point-cloud network of the PointNet type~\cite{Qi2017}. Each particle enters through its three-momentum components; a shared per-particle multilayer perceptron with layer widths $3\rightarrow64\rightarrow64\rightarrow32$ is applied to every particle, and max and mean pooling over the particles produce a 64-dimensional event descriptor. The descriptor is passed through a projector consisting of two dense layers of width 32, and the resulting 32-dimensional vector is normalized to unit length. We denote the embedding of event $a$ by $\mathbf{z}_{a}$.

The network is trained so that the embedding reproduces the geometry of the transport feature space. Within each minibatch, the $k=16$ nearest neighbors of each event in $F$ space are taken as positives, and the InfoNCE loss~\cite{Oord2018}
\begin{equation}
	\mathcal{L}=-\frac{1}{k}\sum_{i}\sum_{j\in\mathcal{N}_{k}(i)}
	\log\frac{\exp\left(\mathbf{z}_{i}\cdot\mathbf{z}_{j}/\tau\right)}
	{\sum_{m\neq i}\exp\left(\mathbf{z}_{i}\cdot\mathbf{z}_{m}/\tau\right)}
	\label{eq:infonce}
\end{equation}
is minimized, where $\mathcal{N}_{k}(i)$ is the set of $k$ positives of event $i$, the dot product of the normalized embeddings is the cosine similarity, $\tau=0.15$ is a temperature parameter, and the sum in the denominator runs over the other events of the batch. In contrast to a direct regression of $F$, the contrastive objective preserves the local neighborhood structure of the transport metric and does not require the network to reproduce the absolute scale of each landmark distance. Training uses the Adam optimizer with a learning rate of $10^{-3}$ and a batch size of 256 for 100 epochs, with a fixed random seed.

Once trained, the network is frozen and applied to all events. The standard $k$-means algorithm produces clusters of unequal size, which would not correspond to fixed fractions of the event sample. We therefore use the $k$-means-constrained algorithm~\cite{Bradley2000, kmeansconstrained}, which makes all clusters contain the same number of events. With $k=20$ classes each class contains 5\% of the events, and the cluster boundaries are determined on the sample itself, in the same way as the percentile calibration of the conventional method. The classes are ordered by their mean $N_{\rm part}$ for labeling and presentation; this is the only place where $N_{\rm part}$ enters, and in a data analysis the same ordering can be obtained from the mean observed $N_{\rm ch}$, which is monotonic in $N_{\rm part}$ inside the acceptance.

\subsection{\label{sec:data}UrQMD simulation}
\begin{table}[!hptb]
	\centering
	\begin{tabular}{@{}c|c|c|c@{}}
		\toprule
		\multicolumn{4}{c}{UrQMD Au+Au collisions}     \\ \midrule
		$\sqrt{s_{\rm NN}}$ / GeV & 3.0  & 5.2  & 19.6 \\ \midrule
		Training events           & 2k   & 2k   & 2k   \\ \midrule
		Inference events          & 7.0M & 6.1M & 5.2M \\ \bottomrule
	\end{tabular}
	\caption{Number of events used for training and inference in Au+Au collisions at $\sqrt{s_{\rm NN}}$ = 3.0, 5.2, and 19.6 GeV from UrQMD model for optimal transport contrastive learning in this work.}
	\label{tab:dataset}
\end{table}
The events are generated with the hadronic transport model UrQMD, version 3.4~\cite{Bass1998,Bleicher1999}, in the default cascade mode, for Au+Au collisions at $\sqrt{s_{\rm NN}}$ = 3.0, 5.2, and 19.6 GeV, with impact parameters sampled within $0\leq b\leq16$ fm. The three energies correspond to the STAR fixed-target (3.0 and 5.2 GeV) and collider (19.6 GeV) configurations. The total numbers of simulated events are 7.04, 6.15, and 5.17 million, respectively. Events without any binary collision are discarded, since no centrality class can be assigned to them.

The first 2000 events with $N_{\rm part}>0$ are used as the training sample, both for the selection of the landmarks and for the training of the network; the remaining 7.0, 6.1, and 5.2 million events are assigned centrality classes by the trained network (Table~\ref{tab:dataset}). The final-state particles of UrQMD are used directly, and the acceptance is the only detector-related ingredient of the analysis.

\section{\label{sec:result}Results}
\begin{figure*}[]
	\centering
	\includegraphics[width=\linewidth]{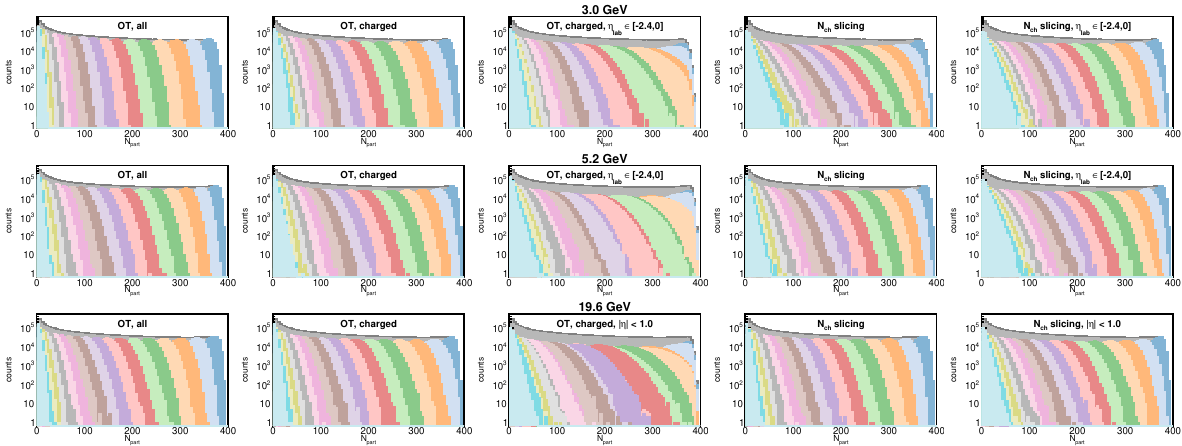}
	\caption{$N_{\rm part}$ distributions in Au+Au collisions at $\sqrt{s_{\rm NN}}$ = 3.0, 5.2, and 19.6 GeV from UrQMD model. The colored shaded areas are $N_{\rm part}$ distributions from 0-5\%, 5-10\%, ..., and 75-80\% centrality bins determined by each method.}
	\label{fig3}
\end{figure*}

In this section, we present the results of centrality determination using the OT-contrastive learning method. The UrQMD events in Au+Au collisions at $\sqrt{s_{\rm NN}}$ = 3.0, 5.2, and 19.6 GeV are used to demonstrate the method. The results are compared with the conventional method of cutting on charged particle multiplicity $N_{\rm ch}$. The notations of results shown in this section are summarized in Table~\ref{tab:notation}. The OT contrastive learning method using all particles is denoted as ``OT, all'', the OT contrastive learning method using only charged particles is denoted as ``OT, charged'', the OT contrastive learning method using only charged particles with $\eta$ cut is denoted as ``OT, charged, $\eta$'', the conventional method of cutting on $N_{\rm ch}$ distribution is denoted as ``$N_{\rm ch}$ slicing'', and the conventional method of cutting on $N_{\rm ch}$ distribution with $\eta$ cut is denoted as ``$N_{\rm ch}$ slicing, $\eta$''. Specifically, the $\eta$ cut is applied to charged particles with $-2.4 < \eta < 0$ in the laboratory frame for $\sqrt{s_{\rm NN}}$ = 3.0 and 5.2 GeV, and $|\eta| < 1$ in the center-of-mass frame for $\sqrt{s_{\rm NN}}$ = 19.6 GeV, which are kinematic cut used in STAR fixed-target and collider data analysis, respectively.

\begin{figure*}[htpb]
	\centering
	\includegraphics[width=\linewidth]{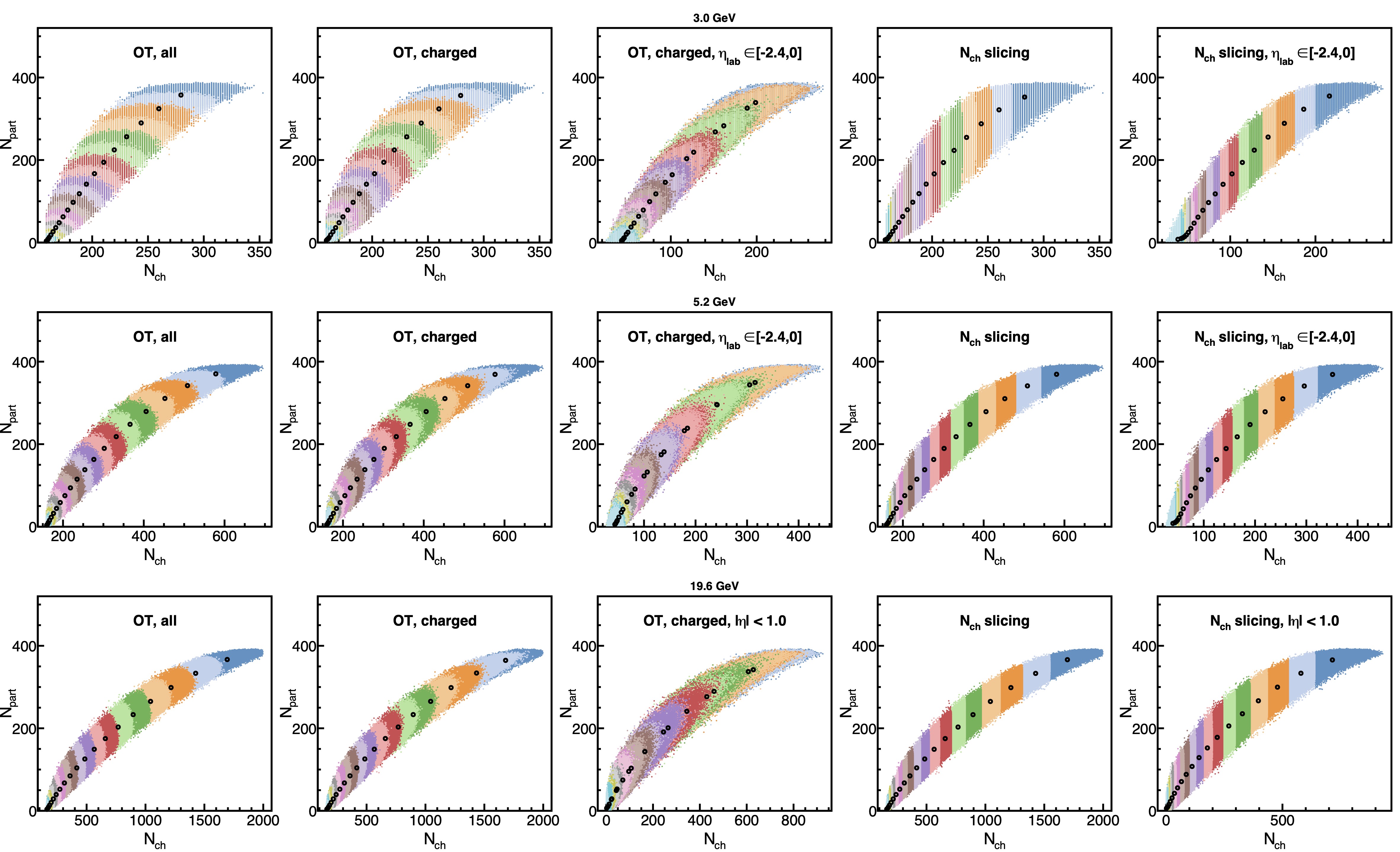}
	\caption{The correlation distributions between $N_{\rm part}$ and $N_{\rm ch}$ in Au+Au collisions at $\sqrt{s_{\rm NN}}$ = 3.0, 5.2, and 19.6 GeV from UrQMD model. The colored shaded areas are $N_{\rm part}$ distributions from 0-5\%, 5-10\%, ..., and 75-80\% centrality bins determined by each method.}
	\label{fig4}
\end{figure*}

\begin{figure*}[htpb]
	\centering
	\includegraphics[width=\linewidth]{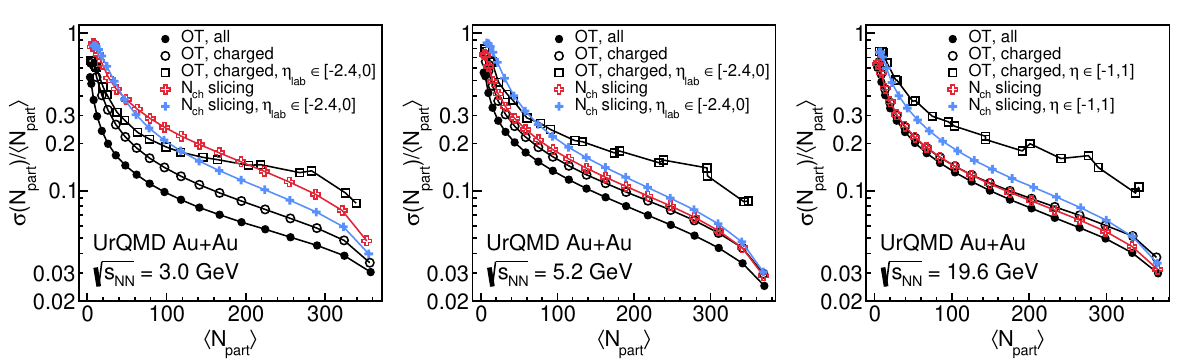}
	\caption{The centrality resolution, defined as the ratio of the standard deviation to the mean of the $N_{\rm part}$ distribution in Au+Au collisions at $\sqrt{s_{\rm NN}}$ = 3.0, 5.2, and 19.6 GeV from UrQMD model.}
	\label{fig5}
\end{figure*}

\begin{figure*}[htpb]
	\centering
	\includegraphics[width=\linewidth]{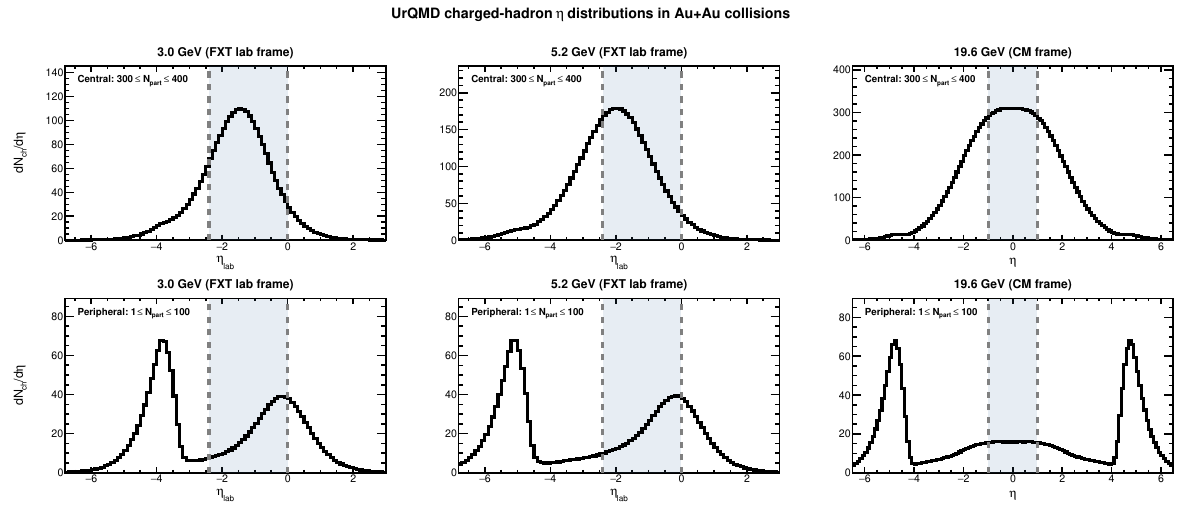}
	\caption{The $\eta$ distributions of charged particles in Au+Au collisions at $\sqrt{s_{\rm NN}}$ = 3.0, 5.2, and 19.6 GeV from UrQMD model for two central ($300 \leq N_{\rm part} \leq 400$) and peripheral ($1 \leq N_{\rm part} < 100$) collisions. The gray dashed lines at 3.0 and 5.2 GeV indicate the acceptance ($\eta$ in lab frame) covered by the STAR fixed-target experiment while the gray dashed lines at 19.6 GeV indicate the acceptance ($\eta$ in center-of-mass frame) covered by the STAR collider experiment.}
	\label{fig6}
\end{figure*}
Figure~\ref{fig3} shows the $N_{\rm part}$ distributions at $\sqrt{s_{\rm NN}}$ = 3.0, 5.2, and 19.6 GeV from UrQMD model obtained with the five methods shown in Tab.~\ref{tab:notation}. The colored shaded areas are $N_{\rm part}$ distributions from 0-5\%, 5-10\%, ..., and 75-80\% centrality bins. It can be seen that the results using OT contrastive learning method (``OT, all'', ``OT, charged'', and ``OT, charged, $\eta$'') can successfully cluster events with similar $N_{\rm part}$ into the same centrality class. The ``OT, all'' which utilize all particles has best performance, while the ''OT, charged, $\eta$'' which utilize only charged particles with a realistic $\eta$ cut in experiments has worst performance and even worse than the conventional method of cutting on $N_{\rm ch}$ distribution (``$N_{\rm ch}$ slicing'' and ``$N_{\rm ch}$ slicing, $\eta$''). The situation is more clearly seen in Fig.~\ref{fig4}, which shows the correlation distributions between $N_{\rm part}$ and $N_{\rm ch}$ at $\sqrt{s_{\rm NN}}$ = 3.0, 5.2, and 19.6 GeV from UrQMD model. For the ``OT, all'' result, events clustered in same centrality class are organized along $N_{\rm part}$ with : events share similar $N_{\rm part}$ but differ substantially in $N_{\rm ch}$. $N_{\rm ch}$ slicing does the reverse: it groups events by charged multiplicity, so a single slice mixes collisions whose $N_{\rm part}$ differ by a large amount. The OT-contrastive method therefore separates events by $N_{\rm part}$, while the slicing separates them by multiplicity. Recall that $N_{\rm part}$ label is never used in the training, only the momentum distributions of final particles as well as $N_{\rm ch}$ are used. The results indicate that the OT-contrastive learning method can extract more information about the collision geometry from final particle distributions. It can also be seen that the OT-contrastive learning method is stable across different collision energies.

As already seen in Fig.~\ref{fig3} that the ``OT, charged, $\eta$'' result yields clearly wider $N_{\rm part}$ distribution per centrality class than the other two ``OT'' cases, the correlation between $N_{\rm part}$ and $N_{\rm ch}$ is also clearly worse than the ``OT, all'' and ``OT, charged'' results, which centrality classes are not well separated - there are overlaps between different centrality classes. But comparing to $N_{\rm ch}$ slicing results, the ``OT, charged, $\eta$'' result still shows that events are grouped according to $N_{\rm part}$ (horizontal bands). This indicates that the OT-contrastive learning method can still extract some information about collision geometry.

We quantify the performance of centrality determination by the centrality resolution, which is defined as the ratio of the standard deviation to the mean ($\frac{\sigma(N_{\rm part})}{\langle N_{\rm part}\rangle}$) of the $N_{\rm part}$ distribution in each centrality class. The smaller $\frac{\sigma(N_{\rm part})}{\langle N_{\rm part}\rangle}$, the better the centrality resolution. Fig.~\ref{fig5} shows the centrality resolution as a function of centrality classes at $\sqrt{s_{\rm NN}}$ = 3.0, 5.2, and 19.6 GeV from UrQMD model. It can be seen that the ``OT, all'' result has the best centrality resolution, while the ``OT, charged, $\eta$'' result has the worst centrality resolution and even worse than the conventional method of cutting on $N_{\rm ch}$ distribution (``$N_{\rm ch}$ slicing'' and ``$N_{\rm ch}$ slicing, $\eta$''). Comparing the two $N_{\rm ch}$ results in this figure, it can be seen that at 3.0 GeV, the $N_{\rm ch}$ slicing with the $\eta$ cut performs better than the version without it. But it is not the case for 5.2 and 19.6 GeV. This could be explained by that, spectator protons make up a large fraction of $N_{\rm ch}$ at 3.0 GeV, so removing the forward region sharpens the correlation between $N_{\rm ch}$ and $N_{\rm part}$.

We now discuss why the $\eta$ cut degrades the performance of the OT method. Figure~\ref{fig6} shows the $\eta$ distributions of charged particles in Au+Au collisions at $\sqrt{s_{\rm NN}}$ = 3.0, 5.2, and 19.6 GeV from UrQMD model for two central ($300 \leq N_{\rm part} \leq 400$) and peripheral ($1 \leq N_{\rm part} < 100$) collisions. The gray dashed lines at 3.0 and 5.2 GeV indicate the acceptance ($\eta$ in lab frame) covered by the STAR fixed-target experiment while the gray dashed lines at 19.6 GeV indicate the acceptance ($\eta$ in center-of-mass frame) covered by the STAR collider experiment. It can be seen that for central collisions, most of charged particles are within the $\eta$ cut, while for peripheral collisions, a large fraction of charged particles are outside the $\eta$ cut. Specifically, there are a large number of spectator protons in peripheral collisions shown at around $\eta_{\rm lab}$ = -4 for 3.0 GeV, $\eta_{\rm lab}$ = -3 for 5.2 GeV, and $\eta$ = $\pm$5 for 19.6 GeV. Inside the acceptance the central--peripheral contrast is therefore carried mostly by the multiplicity itself, which the transport distance deliberately removes; this is why ``OT, charged, $\eta$'' falls below simple multiplicity slicing.

\section{\label{sec:summary}Summary}
In this paper, we proposed an unsupervised method to determine collision centrality of heavy-ion collisions from final particle momentum distributions. The method combines the Wasserstein-1 distances in three-momentum space of each event to a fixed set of landmark events with a PointNet embedding trained by contrastive learning, and partitions the embedding into 20 centrality classes by constrained $k$-means clustering. Training was performed on 2000 events per energy, and inference on 7.0, 6.1, and 5.2 million events, for Au+Au collisions at $\sqrt{s_{\rm NN}} = 3.0$, 5.2, and 19.6 GeV generated with the UrQMD model, respectively. We found that the method, when using all outgoing particles, yields a better centrality resolution than the conventional method of cutting on particle multiplicity commonly used in experiments. However, the method encounters a bottleneck when realistic kinematic cuts are applied to the particles, which could be explained by that the distinct features in momentum distribution between central and peripheral collisions are lost. Further studies are needed to apply optimal transport to centrality determination in heavy-ion collisions.

\section*{Acknowledgement}
Shu-Yun Yang is supported by the Guangxi Young and Mid-Career University Teachers' (Research) Basic Capacity Enhancement Project (No. 2025KY0533). Yu Zhang was supported by the Guangxi Natural Science Foundation (No. 2025GXNSFBA069067).

\bibliography{apssamp}

\end{document}